\documentclass[11pt]{article}
\usepackage[utf8]{inputenc}
\usepackage{listings}
\usepackage{color}
\usepackage{graphicx}
\usepackage{longtable}
\usepackage{float}
\usepackage{amsmath}
\usepackage{amssymb}
\usepackage{authblk}

\title{}
\date{\today}
\begin{document}

\title{Stochastic single-particle based simulations of cellular signaling embedded into computational models of cellular morphology}
\author{Thorsten Pr\"ustel} 
\author{Martin Meier-Schellersheim} 
\affil{Computational Systems Biology Section\\Laboratory of Immune System Biology\\National Institute of Allergy and Infectious Diseases\\National Institutes of Health, Bethesda, Maryland 20892, USA}
\maketitle
\let\oldthefootnote\thefootnote 
\renewcommand{\thefootnote}{\fnsymbol{footnote}} 
\footnotetext[1]{Email: prustelt@niaid.nih.gov, mms@niaid.nih.gov} 
\let\thefootnote\oldthefootnote
\begin{abstract}
Cells exhibit a wide variety of different shapes.
This diversity poses a challenge for computational approaches that attempt to shed light on the role cell geometry plays in regulating cell physiology and behavior.
The simulation platform Simmune is capable of embedding the computational representation of signaling pathways into realistic models of cellular morphology.
However, Simmune's current approach to account for the cell geometry is limited to deterministic models of reaction-diffusion processes, thus providing a coarse-grained description that ignores stochastic local fluctuations.
Here we present an extension of Simmune that removes these limitations by employing an alternative computational representation of cellular geometry that is smooth and grid-free. These features make it possible to incorporate a fully stochastic, spatially resolved description of the cellular biochemistry. The alternative computational representation is compatible with Simmune's current approach for specifying molecular interactions. This means that a modeler using the approach needs to create a model of cellular biochemistry and morphology only once to be able to use it for both, deterministic and stochastic simulations.
\end{abstract}
\section{Introduction}\label{introduction} 
Computational approaches that aim at gaining a mechanistic understanding of cellular behavior by generating realistic predictive models of cellular signaling networks face several challenges. First, signaling pathways dynamically emerge from interactions among many signaling proteins and lipids. These may form multi-molecular complexes, assembling and disassembling over time in a highly organized manner \cite{Jones:2006aa, Brown:2009aa, Hong:2010aa}. Frequently, signal-transducing complexes are characterized by several binding sites and internal states that, in the course of molecular interactions, undergo post-translational modifications, thereby modifying their biochemical behavior and generating new interaction possibilities. This dynamically changing, combinatorial complexity of signaling pathways may result in prohibitively large numbers of possible functional states, rendering conventional approaches that describe a pathway by explicitly describing its mass-action kinetics inapplicable \cite{Hlavacek:2003aa}.

Importantly, cellular signaling networks operate within given cellular morphologies defining spatial contexts that can only be ignored if one assumes that the biochemistry is well-stirred and evolves in a static, simple shaped geometry, such as a cube. However, these assumptions are frequently not justified, as is obvious when looking at the typically heterogeneous spatiotemporal distributions of concentrations within cells \cite{Kholodenko:2003aa, Maeder:2007aa} and the wide variety of intricate cellular morphologies. Because the specific cellular shape can act on the biochemistry and vice versa, computational representations of cellular shape are required that are flexible enough to reflect this diverse biological reality. 

Moreover, having models of cellular biochemistry and computational representations of a cell’s morphology at our disposal, they have to be integrated with each other. This step typically requires adapting algorithms that were developed for simple flat geometries, such as planes and cubes. Given the variety of cellular shapes and signaling pathways that can be combined with each other this adaptation should occur automatically and be left to the computer. An automatic incorporation requires, however, that the computational models of cellular morphology and biochemistry are not ‘hardwired’ but can be freely interchanged. 

The simulation software Simmune addresses these challenges and can be used to generate predictive spatially-resolved models \cite{Angermann:2012aa, Angermann:2019aa}. Simmune is designed and implemented in a highly modular fashion and provides software facilities that are capable of creating computational models of signaling pathways and cellular morphology independently from each other. The Simmune Modeler \cite{Zhang:2013aa} provides a visual language for molecular states and bimolecular interactions that allows experimentalists without computational background to create realistic models of cellular signaling pathways. 
The difficulty to specify ‘upfront’ all possible reactions within a reaction network is avoided by a rule-based approach \cite{Meier-Schellersheim:2006aa, Hlavacek:2006aa, Lok:2005aa, Feret:2009aa, Koschorreck:2008aa, Mallavarapu:2009aa, Faeder:2009aa}: A user only has to specify pairwise interactions among molecular binding sites. The Simmune Network Viewer \cite{Cheng:2014aa} can then visualize how these interactions are interconnected and Simmune automatically generates the full resulting network of all possible multi-molecular complexes. Computational models of cellular geometry can be created with the Simmune CellDesigner. This software tool provides a graphical user interface (GUI) enabling a user to populate the simulated system with spatially discretized versions of cells, their intra-cellular organelles and of the extracellular space. It also permits to assign certain regions to act as the locations of the initial biochemistry. 

Finally, the Simmune Simulator automatically integrates the representations created by the Modeler and CellDesigner and computes the spatially-resolved dynamics as time series and concentration profiles of the system \cite{Angermann:2012aa, Angermann:2019aa}.  
 
Currently, Simmune models the time evolution of the reaction network as a deterministic process  and molecular populations as concentrations. As a consequence, it provides a coarse-grained description in terms of averaged mean-field quantities and ignores stochastic fluctuations. For many cellular signaling systems, e.g. those that involve large molecule copy numbers, a deterministic description is justified and even warranted, because an alternative stochastic description would be prohibitively inefficient without contributing additional insights into the system's behavior. 

However, at a molecular scale, all processes underlying cell physiology are fundamentally stochastic phenomena. Biochemical reactions appear as probabilistic transformations of or interactions between stochastically moving discrete molecules. The timing of individual reactions is nondeterministic and the total number of molecules in a given state evolves in a discrete and stochastic, or 'noisy', fashion. As soon as the involved copy numbers are low and one is interested in phenomena for which the discrete nature of molecules and their location is essential, such as receptor clustering, this stochastic aspect of biological reality can no longer be neglected and a deterministic description fails.

In this manuscript, we present an extension of Simmune that removes its limitation to deterministic simulations. In contrast to spatially discretized grid-based approaches, we employ a computational representation of cellular geometry that is smooth and grid-free. These features make it possible to incorporate a fully stochastic, spatially resolved description of the cellular biochemistry. The extension adheres to Simmune’s modular design: the new computational representation is compatible with Simmune’s current approach, thereby a user needs to create a model of cellular morphology only once to be able to explore its behavior through deterministic or stochastic simulations. 

Moreover, the computational representations of the biochemistry created with the Simmune Modeler can be used for the stochastic simulations as well. Thus, the software now offers a flexible combination of computational tools to study cellular behavior: The 'building blocks' of a computational model, such as the specifications of molecular interactions, representations of cellular geometry and the type of dynamics (deterministic or stochastic) can be freely combined. We expect that this flexibility will facilitate simulation studies comparing differences between deterministic and stochastic dynamics that may emerge in spite of shared biochemical rules and cellular morphology. Finally, we anticipate that the close correspondence between the different computational representations of cellular shapes will pave the way to develop hybrid simulators capable of integrating deterministic with stochastic time evolution. 

We point out that there are other software tools available that offer spatially-resolved and/or particle-based stochastic simulation possibilities, such as, for example \cite {czech2009rapid, Feret:2009aa, arjunan2009new, tolle2010meredys, andrews2010detailed, gruenert2010rule, Sneddon:2011aa, hepburn2012steps, masaru2013cell, blinov2017compartmental}.

\begin{figure}
\includegraphics[scale=0.8]{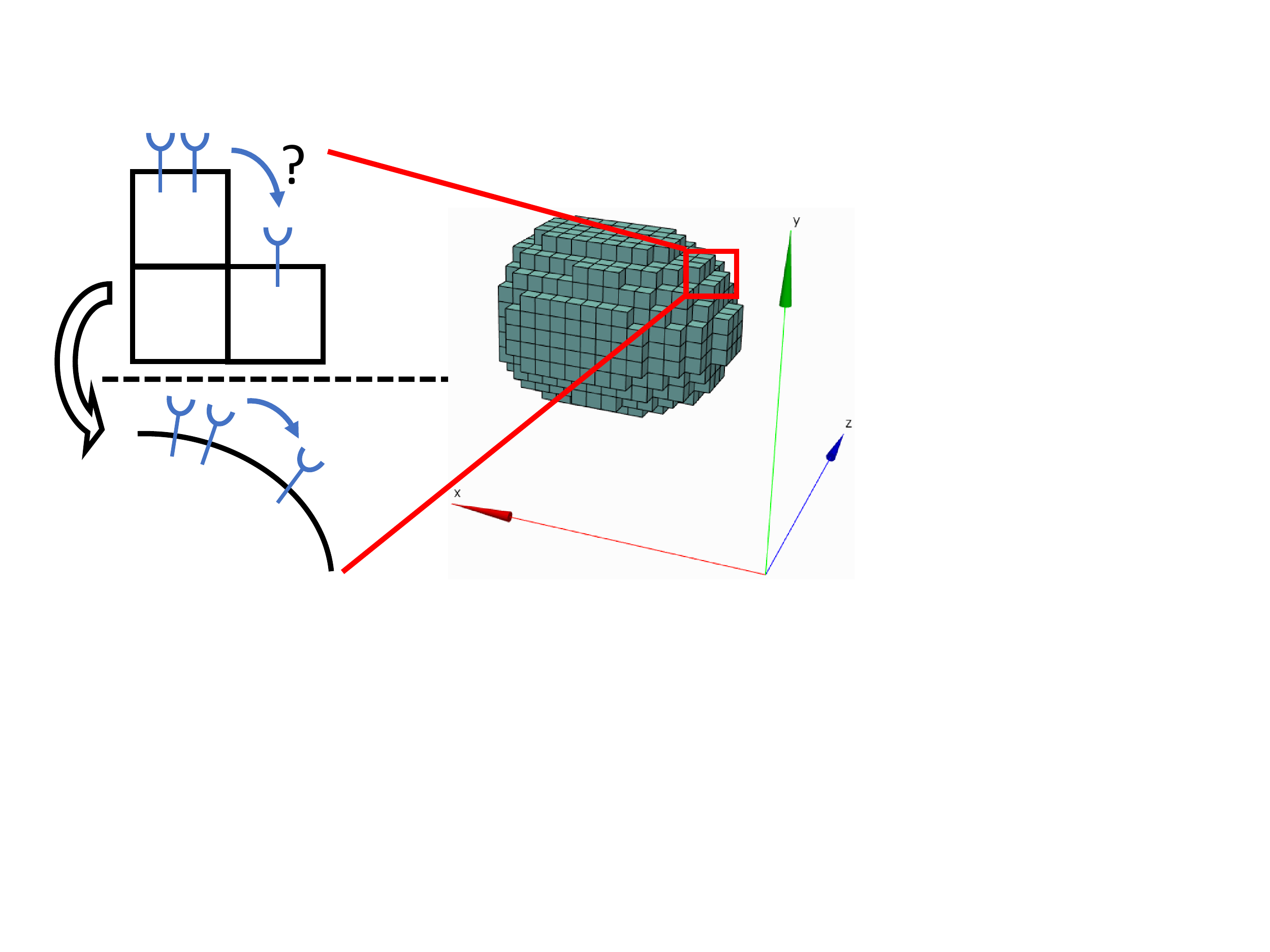}
\caption{
Schematic illustration of the technical difficulties to integrate a discretized grid-like model of cellular morphology as used by Simmune with a stochastic dynamics that generates molecule trajectories.
These difficulties become more severe for surface diffusion, because the stair-like structure would require major modifications of the stochastic dynamics to adequately define Brownian paths around corners and edges.
}\label{fig:diffusion_over_edge}
\end{figure}

\section{Computational models of cellular morphology}  
\label{comp_model_geo}
In Simmune, the geometry of a simulated system consists of the shapes of the simulated cell(s) with their intracellular domains and organelles and the structure of the extracellular space \cite{Angermann:2012aa}. In a deterministic approach, a system of coupled partial differential equations, known as reaction-diffusion equations, governs the time evolution of the spatially resolved biochemistry. Molecular populations are described in terms of local concentrations that are coarse-grained, averaged quantities. Mathematically, concentrations are continuous function of time and space--mathematical idealizations that can only be represented at a finite number of spatial points by a computer. Hence, the numerical solution of the reaction-diffusion equations requires a spatial discretization of the system's geometry. Simmune employs a finite volume scheme \cite{eymard2000finite} that amounts to a uniform division of the simulated space into cubic volume elements, also referred to as voxels. As mentioned in the Introduction, the Simmune CellDesigner and its graphical user interface (GUI) allows users to generate this discretized space and populate it with representations of cells, their membranes, cytosolic compartments and intracellular organelles that are all composed of subsets of voxels.

While the reaction-diffusion equations can be adapted to a grid representation of space in a natural way for cytosolic and extracellular diffusion, this is not longer true for diffusion on the membrane. As discussed in Refs.~\cite{Novak:2007aa, Angermann:2012aa}, the surface elements of the voxels that enclose a simulated cell inadequately approximate the cell’s surface. For instance, the circumference and area of a discretized sphere does not converge to the correct values, even when the grid constant tends to zero. Furthermore, a cubic discretization of the surface Laplacian results in severe artifacts that have to be corrected for by calculating the local membrane curvatures to adapt area, shape and distances between surface elements of the cubic voxels. With these corrections being taken into account, surface diffusion can be accurately described within the finite volume scheme \cite{Novak:2007aa, Angermann:2012aa}.

Whereas cubic discretization artifacts can be controlled for the deterministic case using straightforward numerical corrections, this is no longer the case for stochastic particle-based dynamics. This is because the mathematical formulation of stochastic dynamics \cite{van1992stochastic} is quite different from the corresponding deterministic formulation. The dynamic state of the biochemical system at a point in time is completely specified by the finite set of locations of the individual molecular complexes, rather than by local concentrations. The corresponding equations of motion are stochastic differential equations; their solutions are continuous trajectories (Brownian paths) that describe the stochastic motion of individual molecular complexes in time and space. It is difficult to see how a description based on trajectories could be made compatible with a cubic discretization of space. This becomes especially apparent for the case of surface diffusion, which would require to profoundly adapt the stochastic differential equations in such a way that Brownian paths around corners and edges of cubes can be defined. These technical difficulties together with the above discussed artifacts introduced by the surface discretization in terms of cubic voxels motivate to look for alternative models of geometry.  

Clearly, a smooth grid-free model of the system's geometry is more suitable to support a stochastic dynamics that generates the paths of individual molecular complexes. Furthermore, models based on an explicit parametrization may be excluded, because it would be difficult to construct appropriate parameter maps efficiently for general surfaces. An alternative is offered by implicit surfaces that refer to a widely used class of models, such as blobs, also known as metaballs or soft objects \cite{blinn1982generalization, gomes2009implicit, de2015survey}, for representing smooth objects.
Instead of using an explicit parametrization for the object's surface, implicit surfaces specify certain constraints. For instance, a metaball specifies a scalar function in three dimensional (3d) space
\begin{eqnarray}
\phi &:& \mathbb{R}^{3} \rightarrow \mathbb{R}, \nonumber\\
&& \mathbf{r}=(x, y, z) \rightarrow \phi(\mathbf{r}),
\end{eqnarray}
also known as potential, influence field or density function. In addition, a metaball defines a threshold value $s$ that serves to define the surface $\mathcal{S}$ as isocontour
\begin{equation}
\mathcal{S} = \lbrace \mathbf{r} \in \mathbb{R}^{3} : \phi(\mathbf{r}) = s \rbrace.
\end{equation}
A major advantage of metaballs is  that more complex geometries can readily be generated by a process called blobby modeling \cite{blinn1982generalization} that refers to the smooth blending of several metaballs.
Here, the basic idea is to consider the superposition of the individual influence fields and to define the surface of the resulting object again as the corresponding isocontour
\begin{equation}\label{blobby_blend}
\mathcal{S} = \lbrace \mathbf{r} \in \mathbb{R}^{3} : \sum^{\text{\# metaballs}}_{i}\phi_{i}(\mathbf{r}) = s \rbrace.
\end{equation}

\begin{figure}
\includegraphics[scale=0.5]{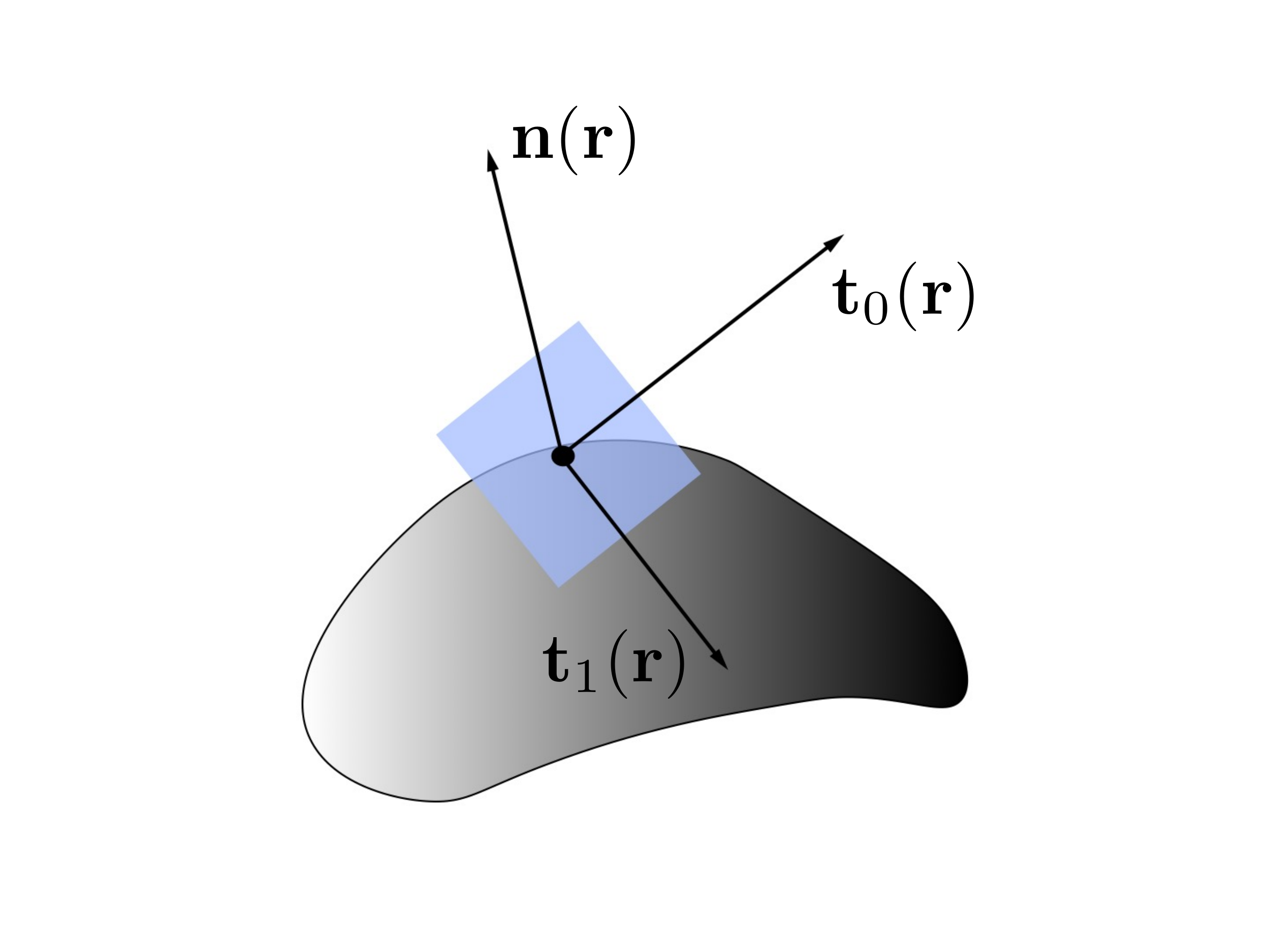}
\caption{
Schematic visualization of the local tangent plane spanned by the tangent vectors $\mathbf{t}_{0}(\mathbf{r})$, $\mathbf{t}_{1}(\mathbf{r})$ and the normal $\mathbf{n}(\mathbf{r})$ at a point $\mathbf{r}$ of the surface of a smooth grid-free represented object. Exploitation of the locally Euclidean structure reduces the degree of technical difficulties encountered in the discrete case, cf.~Fig.~\ref{fig:diffusion_over_edge}. A grid-free representation in terms of compositions of metaballs offers the advantage that the unit normal and, in turn, the local differential geometry such as the tangent plane can readily be determined.  
}\label{fig:local_frame}
\end{figure}

While implicit surfaces can be used directly as computational models, they have also been employed as auxiliary data structures to facilitate the creation of discrete models of geometry, such as surface triangulations \cite{de2015survey, gomes2009implicit}.
In fact, Simmune's generation of a discretized representation of the systems' geometry is implemented in terms of metaballs. In view of possible applications to stochastic simulations, it is now natural to promote these purely auxiliary data structures to the actual computational representation of cellular geometry to enable stochastic simulations. With this extension, a user needs to create a model of the system's geometry only once and obtains automatically both a discretized grid and smooth grid-free representation amenable to deterministic and stochastic simulations. In addition, the close relation between the two representations facilitates the comparison between deterministic and stochastic simulation results.

Finally, let us point out the explicit mathematical form of the potential used by Simmune's metaball system
\begin{equation}\label{potential}
\phi(\mathbf{r}) = \left\{ 
\begin{array}{ll}
 & [1-(r/R)^2)]^{2} \qquad \mathrm{if}\, r< R\\ & 0 \qquad \qquad \qquad \quad \mathrm{otherwise}, 
 \end{array} \right.
\end{equation}
where $r=\vert\mathbf{r}\vert$ and $R$ refers to the metaball's radius.
\section{Integration of computational models of geometry and stochastic integrators}
Simmune uses a particle-based stochastic simulation algorithm \cite{Prustel-sim-2015, prustel2017unified} that is based on the Smoluchowski-Collins-Kimball (SCK) model \cite{smoluchowski:1917, Goesele:1984, Rice:1985} of diffusion-influenced bimolecular reactions.

Different theoretical approaches exist to implement the SCK model's account for the diffusive component of bimolecular reactions \cite{van1992stochastic, gardiner2009stochastic}. The Langevin approach, also known as Brownian Dynamics (BD) in a numerical context, accounts for the time evolution of the molecules' positions directly as specific solutions (trajectories) of stochastic differential equations. In contrast, the Smoluchowski and Einstein diffusion equations govern the time evolution of the conditional probability density function (PDF) $p(\mathbf{r}, t \vert \mathbf{r}_{0})$, also known as Green's function (GF) that provides the probability of finding a molecule at a position $\mathbf{r}$ at time $t$, given it was sharply localized at  $\mathbf{r}_{0}$ at time $t_{0}=0$. 

The SCK framework is widely used in stochastic simulation algorithms to quantify the impact of diffusion on chemical reaction rates \cite{edelstein1993brownian, edelstein1997brownian, kim1999dynamic, barenbrug2002accurate, Zon:2005aa, oppelstrup2009first, Johnson:2014aa, sokolowski2019egfrd}. Frequently, these type of algorithms employ GF to address the problem of the high computational cost that single-particle stochastic models based solely on BD typically entail.

The integration of computational representations of the cellular morphology and the stochastic integrator that we will describe in the following does not depend on the details of the stochastic approach. This 'loose coupling' offers the advantage that different stochastic integrators, even those that do not use the SCK picture, such as Doi-based \cite{Doi_1:1976, Doi_2:1976, Wilemski:1973, Isaacson:2014, prustel2017unified} approaches, could be readily swapped with each other.   

Typically, particle-based stochastic simulation algorithms assume that flat Euclidean spaces, such as planes and cubes, describe the simulated system's geometry and tacitly employ special geometric features only provided by Euclidean structures, such as the existence of a global coordinate system and distance metric. Therefore, the transition to more general geometries motivated by the wish for realistic models of cellular morphology necessitates modifications to the stochastic description. The parts of the algorithms describing reactions that involve a spatial aspect, such as bimolecular associations and unimolecular decays (through the need to place the decay products correctly) have to be adapted, while descriptions of 'non-spatial' reactions of molecular complexes such as phosphorylations remain unaffected. Among the different types of bimolecular reactions, such as associations between two cytosolic complexes and between cytosolic and membrane complexes, the most profound adaptions are necessary for associations between membrane complexes. 

\subsection{Stochastic description of reaction-diffusion systems on curved membranes}\label{stochastic_surface}
Already in the Euclidean case, reaction-diffusion systems in 2d are quite different from corresponding 3d systems, due to distinct features that diffusion exhibits in 2d.  
Two is the critical dimension with regard to recurrence and transience of Brownian walks \cite{Toussaint_Wilczek_1983}, the steady-state solution of the diffusion equation is
inconsistent with the boundary condition at infinity \cite{Emeis_Fehder_1970}, and the time-dependent reaction rate vanishes in the long-time limit. These properties are related to each other and make it more complicated to derive a mass-action rate from the underlying microscopic stochastic description \cite{Yogurtcu:2015aa}.    
In addition, the mathematical treatment appears to be more involved than in 1d and 3d \cite{TPMMS_2012JCP, Grebenkov:2019aa}.
Finally, the 2d case is of particular importance in cell biological applications, being a prerequisite for a quantitative understanding of processes such as signal-induced inhomogeneities and receptor clustering on cell membranes \cite{bethani:2010}, where the receptors' slow lateral diffusion represents a less efficient mixing mechanism than in the three dimensional cytoplasm.

The task at hand is now to incorporate the stochastic algorithms originally developed for Euclidean planes with the computational representations of general cellular shapes, thereby including curvature as a quantity that takes part in the stochastic dynamics. 

In mathematics, the corresponding problem is solved by exploiting the locally flat structure of Riemannian manifolds: at any point $\mathbf{r}_{0}$ of its surface a Riemann manifold gives rise to a local tangent plane, also known as frame, that comes equipped with a locally Euclidean structure \cite{ikeda1989stochastic}. Equations that govern the dynamics in a globally flat space keep their validity in the local frame in a vicinity of $\mathbf{r}_{0}$. The size of surface patch around $\mathbf{r}_{0}$ that is well approximated by the local frame is determined by the local curvature. 

Following this mathematical lead, we now seek as a first step an algorithmic description of free diffusion, i.e., Brownian motion on the surface of a metaball or of a smooth blend of several of them. 

As discussed in Sec.~\ref{comp_model_geo}, this task is equivalent to find a numerical approach to free Brownian motion on a surface defined by the general equation
\begin{equation}\label{implicit_surface}
\phi(\mathbf{r}) = 0,
\end{equation}
see Eq.~\ref{blobby_blend}.
This problem has been solved previously \cite{Hoyst:1999aa} in the spirit of the mathematical constructions described above. Now, while Ref.~\cite{Hoyst:1999aa} considers free diffusion on a general surface, we use the same framework to describe 2d-2d interactions, such as (irr)reversible associations of membrane molecules, and 2d-3d interactions, such as binding reactions between membrane and extracellular or cytosolic complexes, for which the surface and bulk are described by general shapes that can be represented by metaballs. 
\begin{figure}
\includegraphics[scale=.405]{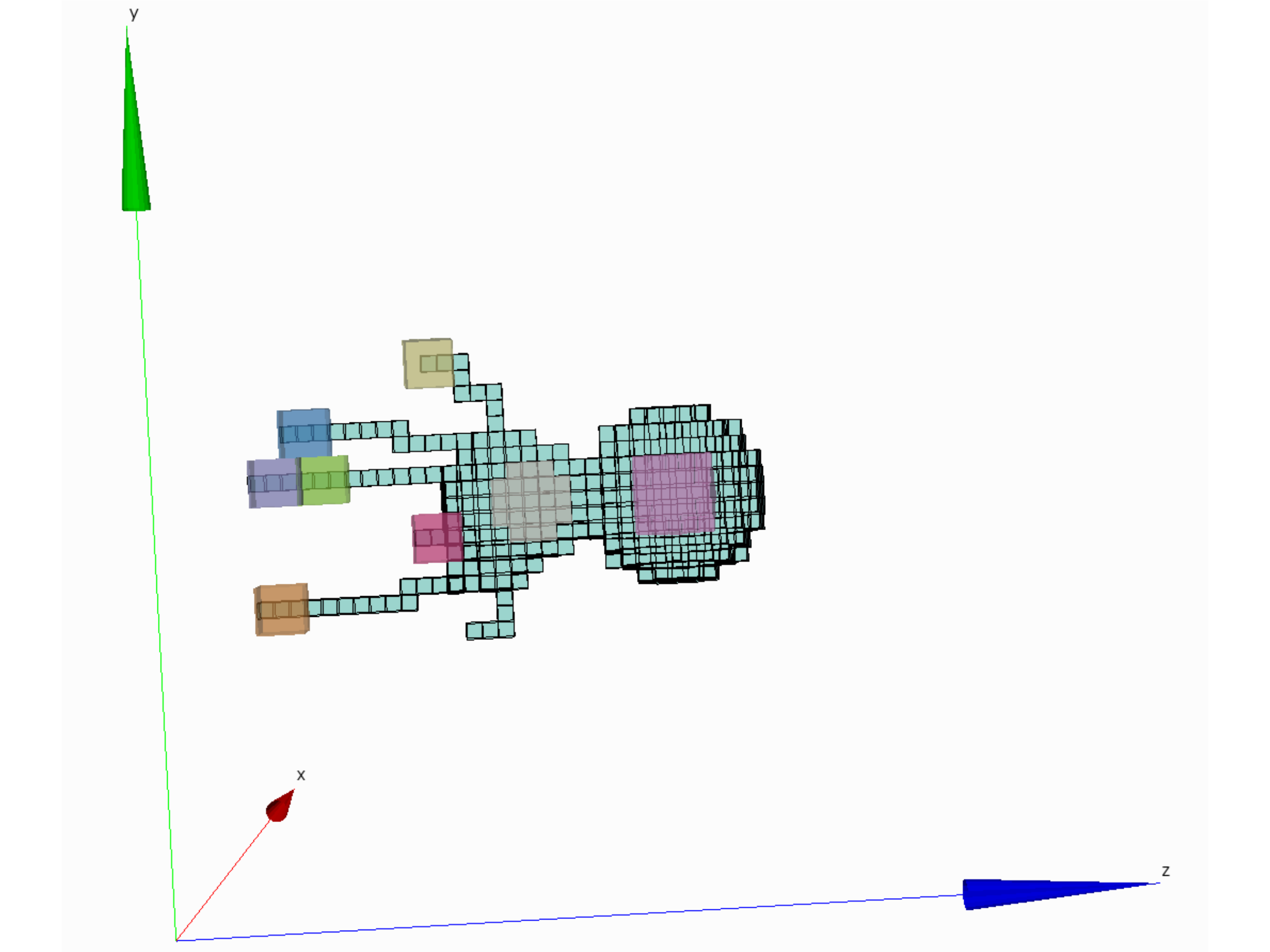}
\includegraphics[scale=.405]{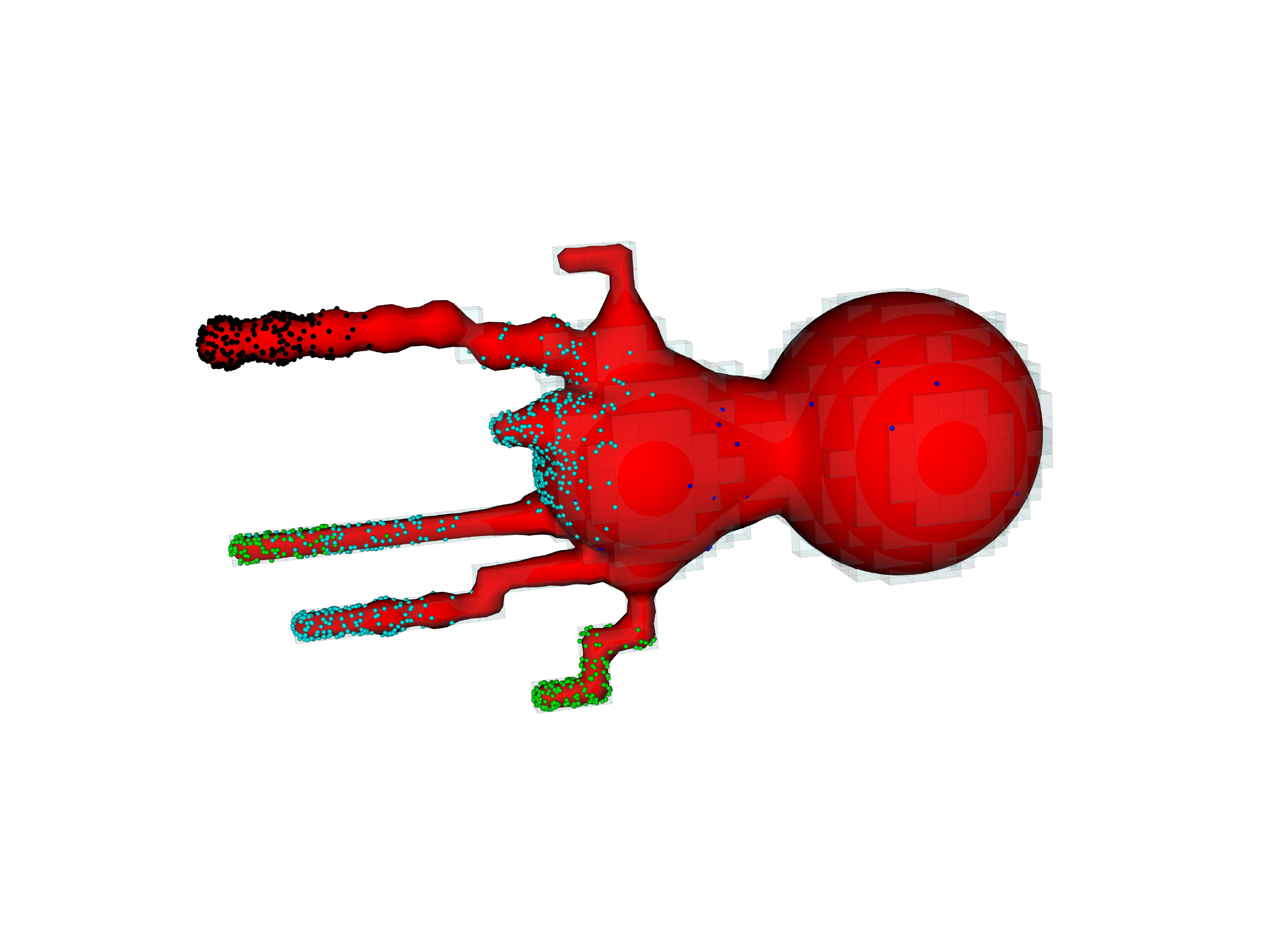}
\caption{
Discrete and smooth representations of cellular morphology. \newline \textbf{Top} A discretized finite volume representation suitable for deterministic simulations created via the Simmune CellDesigner. Colored patches specify regions that allow to define the spatial distribution of the initial biochemistry. \newline \textbf{Bottom} The automatically generated corresponding grid-free, smooth representation of the same cellular morphology amenable to stochastic simulations. For illustration, the grid is also displayed.
}\label{fig:discrete_vs_smooth}
\end{figure}
Following Ref.~\cite{Hoyst:1999aa} and defining the surface as the points that satisfy Eq.~\ref{implicit_surface}, one may characterize the local tangent plane at a point $\mathbf{r}_{0}=(x_{0}, y_{0}, z_{0})$ as the set of points $\mathbf{r}=(x, y, z)$ that satisfy
\begin{equation}
(\mathbf{r} - \mathbf{r}_{0}) \cdot \nabla\phi(\mathbf{r}_{0}) = 0,
\end{equation}
that is, the vector $\mathbf{r} - \mathbf{r}_{0}$ is perpendicular to the gradient $\nabla\phi(\mathbf{r}_{0})$, which is the surface's normal at point $\mathbf{r}_{0}$.

Now, we seek to construct an orthonormal basis of local tangent plane, that is, two mutually perpendicular vectors $\mathbf{t}_{0}$, $\mathbf{t}_{1}$ that span the tangent space at $\mathbf{r}_{0}$.

Knowledge of this basis enables one to define a free Brownian position update analogously to the Euclidean plane case
\begin{equation}\label{free_membrane_propagation}
\mathbf{r}^{\ast}(t+\Delta t) = \mathbf{r}(t)+ \sqrt{2D\Delta t} \xi \mathbf{t}_{0} + \sqrt{2D\Delta t} \chi \mathbf{t}_{1} 
\end{equation}
where $\xi$ and $\chi$ refer to samples from a Gaussian distribution with vanishing mean and variance equal to unity.  
However, the updated position given by Eq.~\ref{free_membrane_propagation} may not fulfill condition Eq.~(\ref{implicit_surface}). Therefore, $\mathbf{r}^{\ast}$ has to be projected back onto the surface as follows \cite{Hoyst:1999aa}
\begin{equation}\label{surface_projection}
\mathbf{r}(t+\Delta t) = \mathbf{r}^{\ast}(t+\Delta t) - \frac{\phi(\mathbf{r}^{\ast}(t+\Delta t))\nabla\phi(\mathbf{r}^{\ast}(t+\Delta t))}{\vert\nabla \phi(\mathbf{r}^{\ast}(t+\Delta t))\vert^{2}}. 
\end{equation}
Here, $\mathbf{r}(t+\Delta t)$ is the final location on the surface after one completed simulation step.

It has been shown \cite{Hoyst:1999aa} that the algorithm represented by Eqs.~(\ref{free_membrane_propagation}), (\ref{surface_projection}) is stable and satisfies the detailed balance condition, provided that $\vert \mathbf{r}(t+\Delta t) - \mathbf{r}^{\ast}(t+\Delta t)\vert$ is much smaller than the local radius of curvature.

Therefore, the local curvature enforces an additional upper limit to the size of the time step $\Delta t$; already in the Euclidean case there are upper limits determined by the requirement that every molecule encounters maximal one other molecule per time step on average \cite{Johnson:2014aa} and  by the unimolecular reactions' time scales.

A major advantage of the metaball representation is that it permits an efficient and straightforward implementation of the algorithm just described. The defining properties of metaballs are critical for a swift construction of the local differential geometry, that is, the local tangent plane and the curvature. Because of Eqs.~(\ref{blobby_blend}), (\ref{implicit_surface}), one can immediately obtain the unit normal vector at any point of the metaball's surface
\begin{equation}\label{unit_normal}
\mathbf{n}(\mathbf{r}):= \nabla\phi(\mathbf{r}) / \vert\nabla\phi(\mathbf{r}) \vert
\end{equation}  
Next, one can exploit that knowledge of $\mathbf{n}(\mathbf{r})$ is sufficient to obtain the tangent plane, see \cite{hughes1999building, frisvad2012building} for algorithms that construct a complete orthonormal basis, given only a single vector. Also the Gaussian curvature can be calculated from the unit normal \cite{spivak1975comprehensive, Hoyst:1999aa}
\begin{equation}
K(\mathbf{r}) = \frac{1}{R_{1}R_{2}}=\frac{1}{2}[-(\partial_{i}\phi_{j}(\mathbf{r}))^{2} + (\nabla\cdot\mathbf{n}(\mathbf{r}))^{2}].
\end{equation}
Here, $R_{1}$ and $R_{2}$ denote two principal radii of the curvature.
In the light of the central importance of the gradient, the simple mathematical form of the potential Eq.~(\ref{potential}) offers an additional advantage: even more complex surfaces composed of several metaballs are represented by a linear superposition of these potentials and hence the resulting gradient itself can be calculated efficiently without the need to resort to numerical differentiation.  

In this way, the construction of a local coordinate system opens up the possibility to reuse Euclidean geometry stochastic algorithms that not only can handle free diffusion but also bimolecular reactions. All subsequent steps of these algorithms, such as detection of encounters/reactions between two molecules and correct two-body propagation of a non-reacting molecule pair, can be extended to models involving general cellular geometries.  
\subsection{Associations between surface and bulk complexes}
As mentioned, the ability to swiftly construct the local differential geometry at any point of a metaball permits to export also techniques that deal with 2d-3d membrane-bulk binding reactions, such as associations between membrane and cytosolic molecules, to general cellular morphologies specified by Eq.~(\ref{implicit_surface}). 

Approaches to stochastic algorithms based on Green's function assisted Brownian Dynamics involve the key step to rewrite the two-body diffusion equation describing an isolated pair of reacting molecules into two diffusion equations \cite{Zon:2005aa, Johnson:2014aa, sokolowski2019egfrd}. The resulting equations describe free diffusion of the center-of-diffusion vector and diffusion of the inter-molecule vector subject to a boundary condition that implements the reaction.

This alternative form of the two-body diffusion equation is enabled by a transformation to a center-of-diffusion coordinate system. While this transformation is straightforward for a pair of two bulk molecules, it becomes more complicated for a pair of one molecule diffusing on a plane and the other moving in the bulk. 

In this scenario, the transformation that separates the two-body diffusion equation reads \cite{sokolowski2019egfrd}
\begin{equation}
\mathbf{r}=
\begin{pmatrix}
x_{3d} - x_{2d}\\
y_{3d} - y_{2d}\\
\epsilon ( z_{3d} - z_{2d})
\end{pmatrix}, \quad \epsilon=\sqrt{1+\frac{D_{2d}}{D_{3d}}}
\end{equation}
and 
\begin{equation}
\mathbf{R}=\frac{1}{D_{2d} + D_{3d}}
\begin{pmatrix}
D_{2d}x_{3d} + D_{3d}x_{2d}\\
D_{2d}y_{3d} + D_{3d}y_{2d}\\
z_{2d}
\end{pmatrix},
\end{equation} provided the 2d plane is described by $z=\text{const}$. Here, $\mathbf{r}_{3d}=(x_{3d}, y_{3d}, z_{3d})$, $D_{3d}$ and $\mathbf{r}_{2d}=(x_{2d}, y_{2d}, z_{2d})$, $D_{2d}$ refer to the position and diffusion constant describing the bulk and surface molecule, respectively.
Now, using the construction of the local differential geometry as described in Sec.~\ref{stochastic_surface}, the center-of-diffusion transformation and all subsequent steps of the stochastic algorithm can be applied with respect to the local frame. In this way, 2d-3d binding reactions can be extended to general models of cellular morphology.   
\section{Numerical Tests}
We performed several tests of the presented simulation algorithm.

\begin{figure}
\includegraphics[scale=0.5]{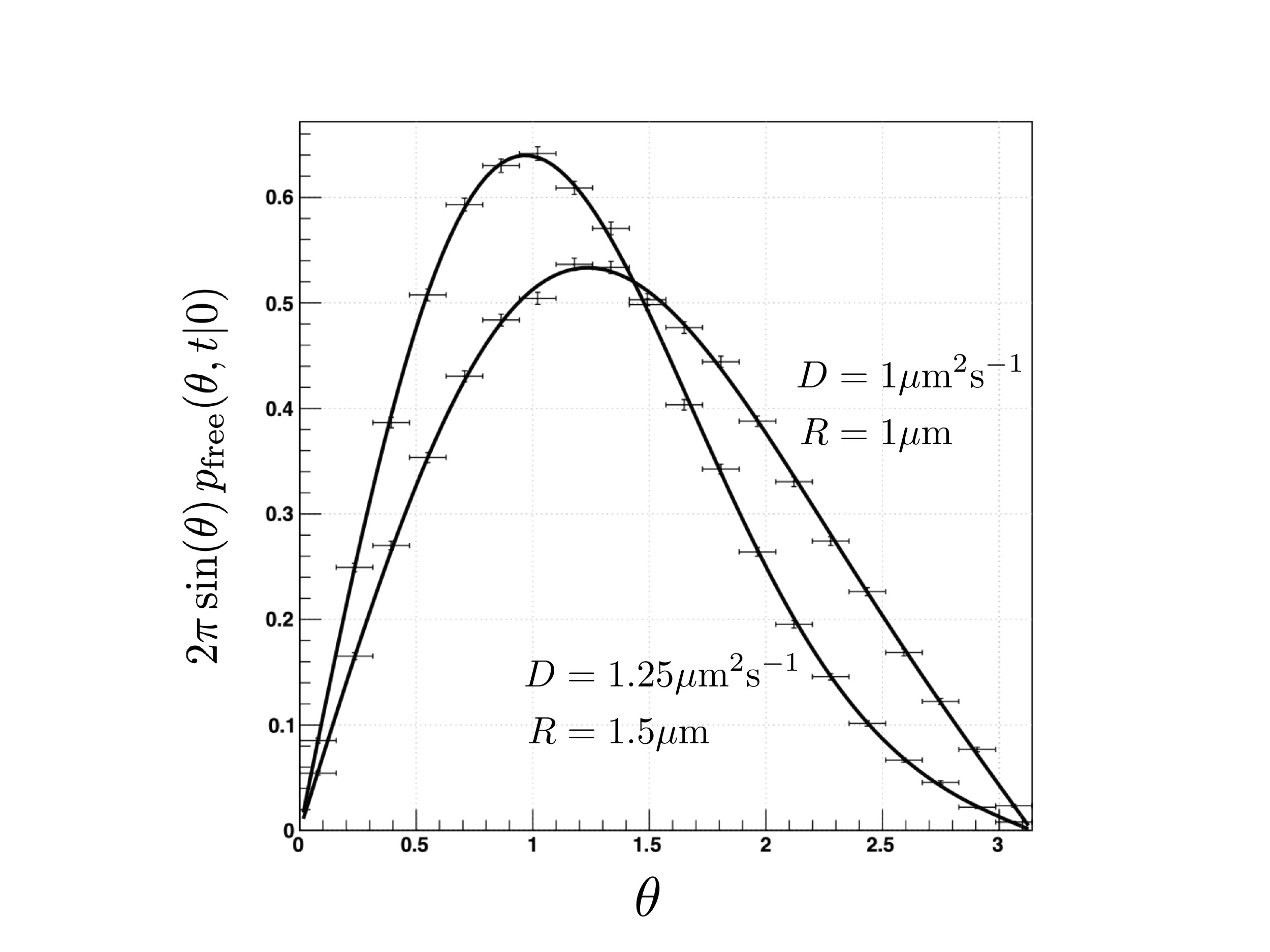}
\caption{
Numerical construction of the GF describing the free diffusive motion of a molecule on the surface of a sphere. 
The figure shows the $\theta$ dependence of the GF for two sets of the diffusion constant and sphere radius $R$, as indicated, at time $t_{\text{sim}}=1\text{s}$.
The solid lines correspond to the exact analytical expression (Eq.~(\ref{p_free_sphere})).
The various markers refer to the stochastic simulation results.
}\label{fig:p_sphere_free}
\end{figure}

First, we simulated free diffusion of a single molecule and the diffusion-controlled irreversible reaction of an isolated pair on the surface of a sphere. The aim was to numerically construct
the corresponding probability density function, i.e., the Green's function $p(\theta, t\vert \theta_{0})$ \cite{Grebenkov:2019aa}, where $\theta$ denotes the polar angle of standard spherical coordinates.
The simulation set up was as follows.
For free diffusion, the molecule was initially placed at the North Pole of the sphere, that is, $\theta_{0}=0$. 
The simulation was run for a time $t_{\text{sim}} =1\text{s}$ during which molecule underwent a diffusive motion and the simulation time step is $\Delta t = 0.001\text{s}$. The final position at time $t_{\text{sim}}$ was recorded and a new simulation was initiated. Fig. \ref{fig:p_sphere_free} shows the simulation results for two different sets of the diffusion constant $D$ and of the sphere radius $R$. The used sets are 
$D=1\mu\text{m}^{2}\text{s}^{-1}$, $R=1\mu\text{m}$ and $D=1.25\mu\text{m}^{2}\text{s}^{-1}$, $R=1.5\mu\text{m}$.
We found excellent agreement between the simulation results and the exact analytical expression of the GF describing free diffusion on a sphere (Eq.~\ref{p_free_sphere}).

\begin{figure}
\includegraphics[scale=0.5]{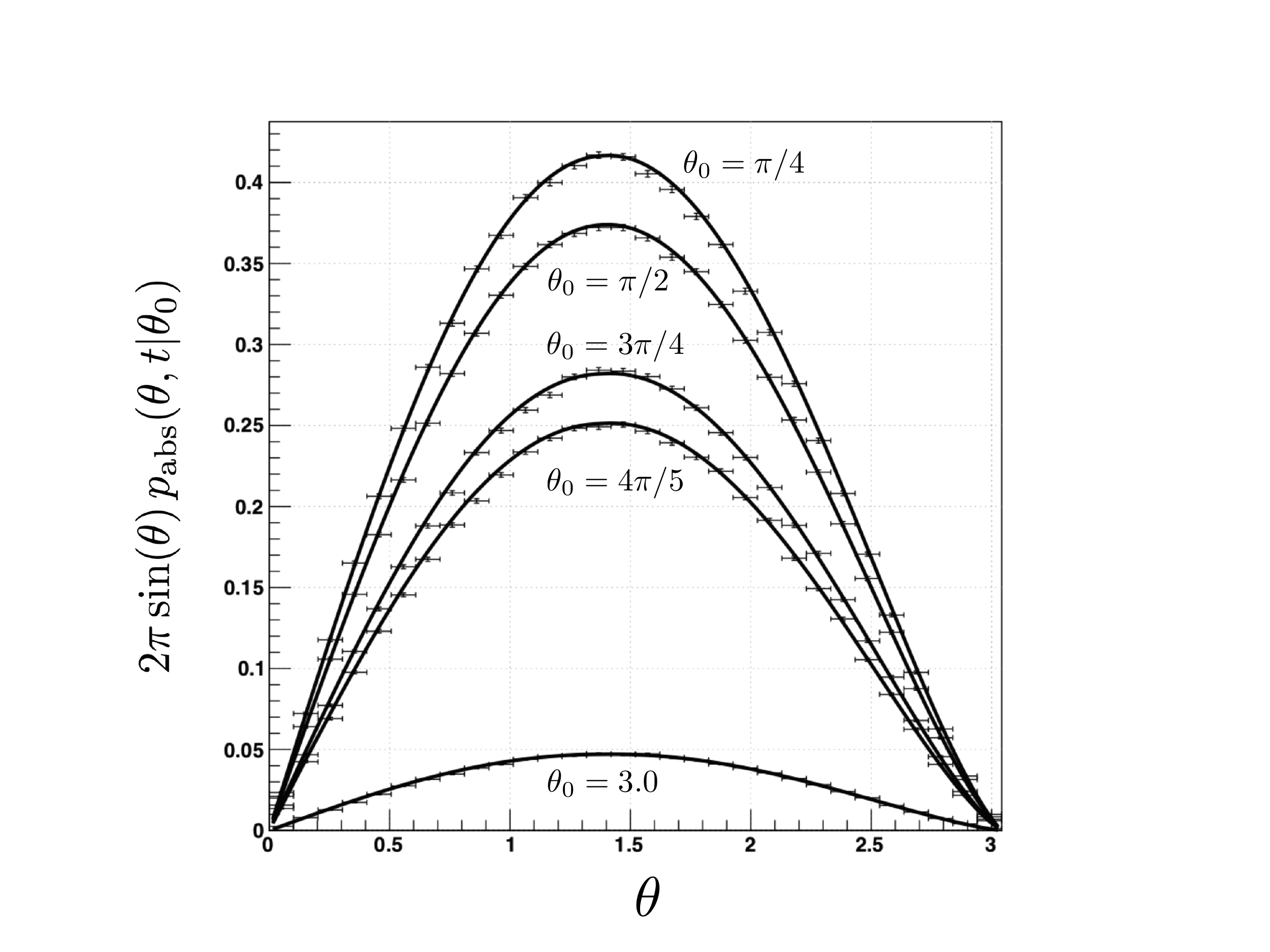}
\caption{
Numerical construction of the GF describing the diffusion-controlled irreversible binding of an isolated pair on the surface of a sphere. 
The figure shows the $\theta$ dependence of the GF satisfying a completely absorbing boundary condition for different values of the initial position $\theta_{0}$, as indicated, at time $t_{\text{sim}}=2\text{s}$. The solid lines correspond
to the exact analytical expressions (Eq.~(\ref{p_abs_sphere})).
The various markers refer to simulation results.
}\label{fig:p_sphere_abs}
\end{figure}

In addition, we considered an isolated pair of molecules $A$ and $B$. Molecule $A$ was held fixed at the sphere's South Pole $\theta = \pi$, while molecule $B$ was placed at several initial positions. The simulation was run for a time $t_{\text{sim}}=2\text{s}$, during which molecule $B$ diffused and potentially associated with molecule $A$ that was modeled as a perfect sink. The radius of the sphere, the effective encounter radius \cite{agmon1990theory, Grebenkov:2019aa} and the diffusion constant are $R=1\mu\text{m}$, $a_{\text{eff}}=0.1\mu\text{m}$ and $D=1\mu\text{m}^{2}\text{s}^{-1}$, respectively.
After each run, we recorded $B$'s final position, unless it was bound to $A$. 
The resulting histogram was normalized to account for the number of bound states at $t_{\text{sim}}$. Fig.~\ref{fig:p_sphere_abs} shows the results for various initial positions together with the corresponding graphs of the exact analytical representation (Eq. (\ref{p_abs_sphere}), see Refs.~\cite{chao1981localization, Grebenkov:2019aa}). Again, we found excellent agreement.

\begin{figure}
\includegraphics[scale=0.5]{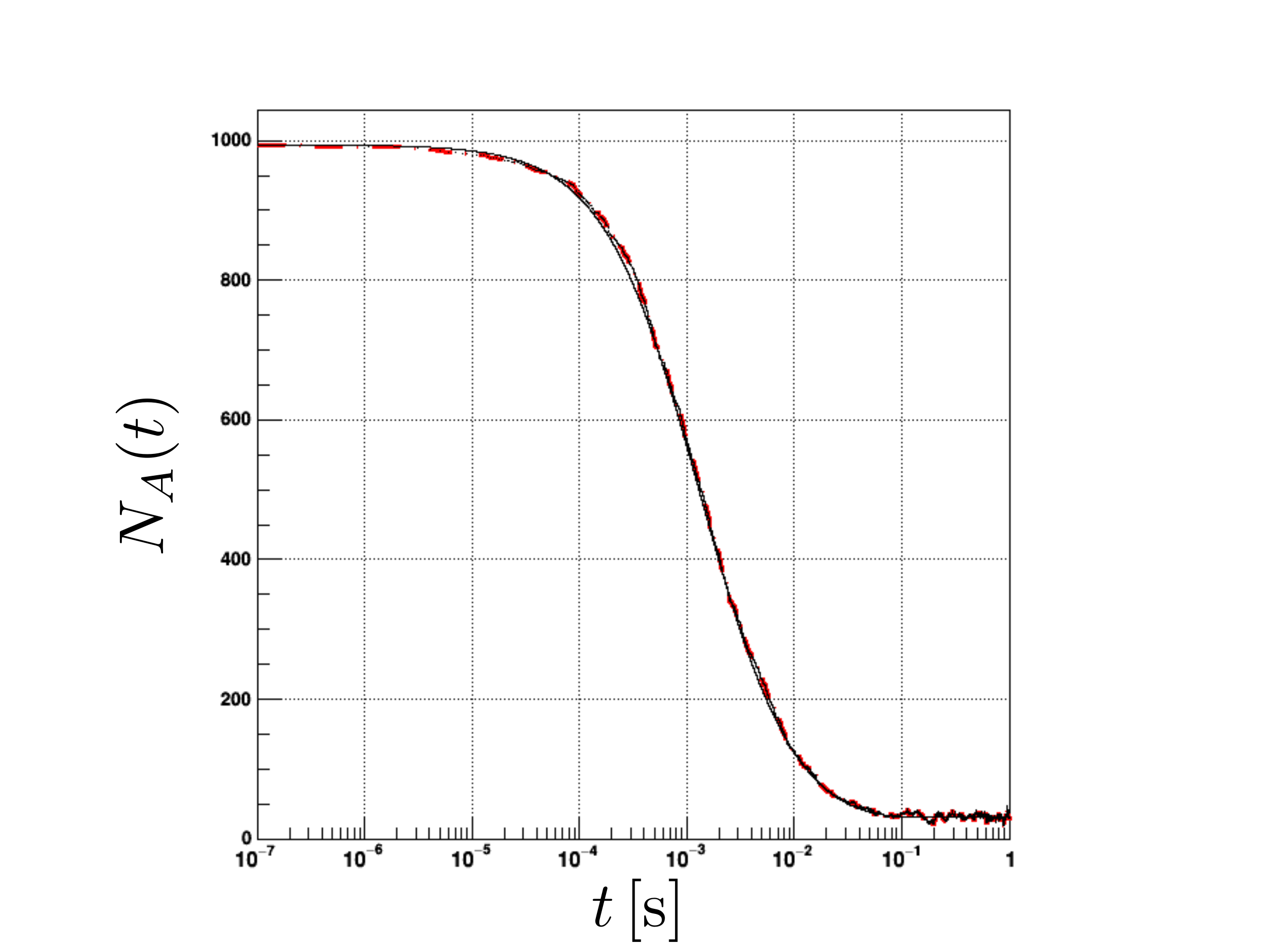}
\caption{
Simulation results vs.~MRE solution for the reversible $A+B\leftrightarrow C$ reaction taking place on the surface of a sphere.
Simulation results and numerical MRE solution are shown in dashed red and solid black curves, respectively, for $D_{A}=D_{B}=1\mu  \text{m}^{2} \text{s}^{-1}$ and
$\kappa_{a}=1\mu \text{m}^{2} \text{s}^{-1}$ and $\kappa_{d}=1 \text{s}^{-1}$.
}\label{fig:rev_2d}
\end{figure}

Next, we turned to many-molecule systems and simulated the reversible binding reaction $A+B\leftrightarrow C$ on a surface of a sphere with surface area $A=0.994 \mu \text{m}^{2}$.
Initially, $N_{A}(t=0)=N_{B}(t=0)=994$ copies of $A$ and $B$ were uniformly distributed over the surface. The simulation was performed with $D_{\text{eff}}=D_{A}+D_{B}=2\mu  \text{m}^{2} \text{s}^{-1}$ and $a_{\text{eff}}=a_{A} + a_{B} = 10^{-3}\mu\text{m}$. Using the dimensionless time step $\Delta \tau = \Delta t D_{\text{eff}} / a^{2}_{\text{eff}} = 0.2$., the simulation was executed until $t_{\text{sim}}= 1\text{s}$ and the copy numbers $N_{A}(t)$, $N_{B}(t)$, $N_{C}(t)$ over time $t$ were recorded.
The further parameters are $\kappa_{a}=1\mu \text{m}^{2} \text{s}^{-1}$ and $\kappa_{d}=1 \text{s}^{-1}$, where $\kappa_{a}$ and $\kappa_{d}$ refer to the intrinsic association and dissociation constant, respectively, which characterize the so-called backreaction boundary condition (BC) \cite{agmon1990theory}. The backreaction BC generalizes the classic Smoluchowski BC to account for diffusion-influenced reversible reactions \cite{Rice:1985, agmon1990theory}. 
Fig.~\ref{fig:rev_2d} shows the simulation result and the numerical solution of the modified rate equation (MRE) approach \cite{szabo1991theoretical, Yogurtcu:2015aa}.

\begin{figure}
\includegraphics[scale=0.5]{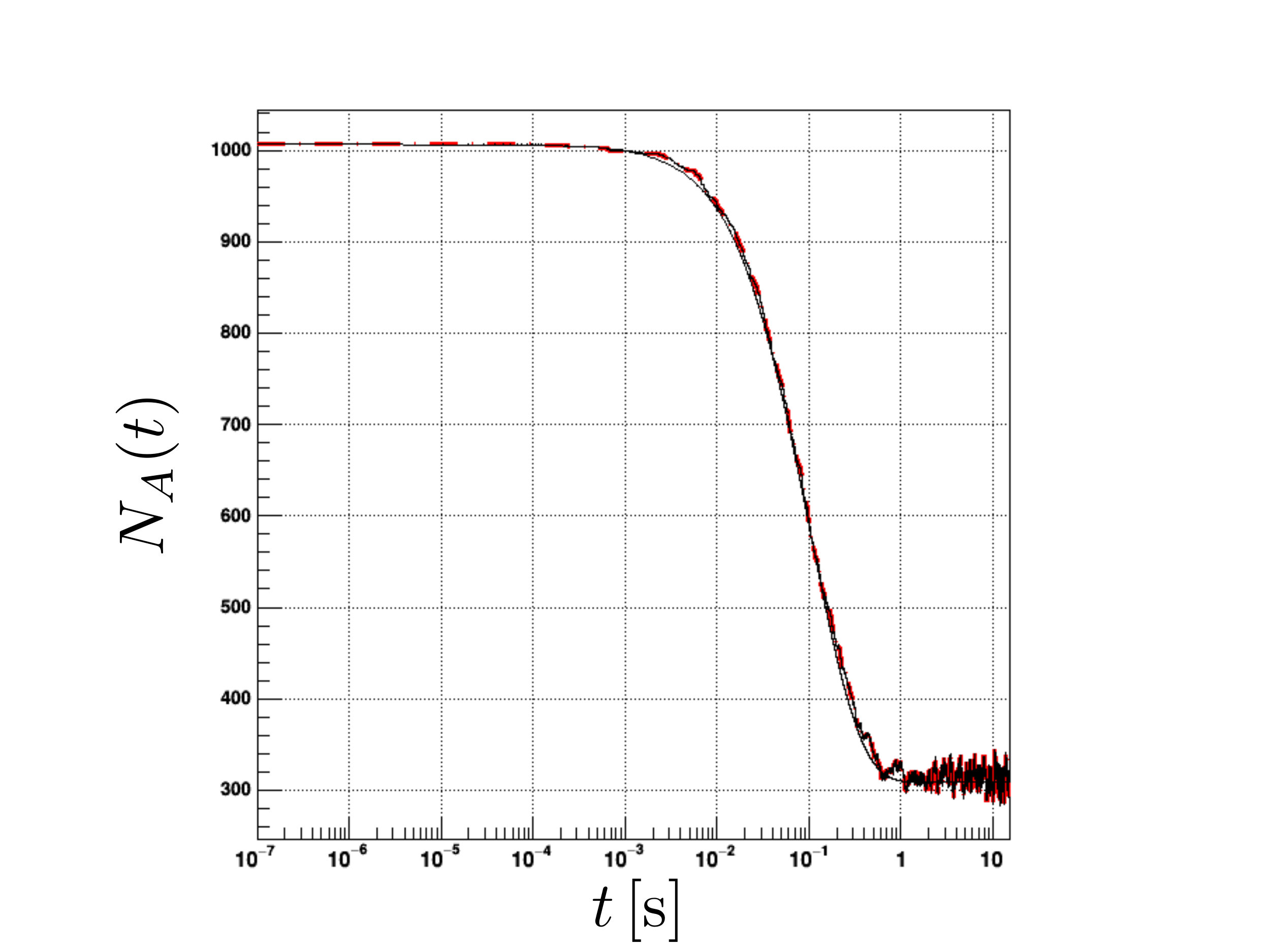}
\caption{
Simulation results vs.~numerical solution of mass-action kinetics for the reversible $A+B\leftrightarrow C$ reaction taking place between $A$ and $B$ molecules, diffusing inside the sphere and on its surface, respectively. Simulation results and numerical solutions are shown in dashed red and solid black for $D_{A}/10=D_{B}=1\mu  \text{m}^{2} s^{-1}$ and
$\kappa_{a}=5\cdot 10^{-4}\mu  \text{m}^{3} s^{-1}$ and $\kappa_{d}=1 s^{-1}$.
}\label{fig:rev_2d_3d}
\end{figure}

Finally, we considered reversible associations between membrane and cytosolic molecules. To this end, we simulated the reversible binding reaction $A+B\leftrightarrow C$, initially uniformly placing $N_{A}(t=0)=1007$, $N_{B}(t=0)=991$ copies inside a sphere with radius $R=0.249\mu\text{m}$ and on its surface, respectively. 
The further parameters are $\kappa_{a}=5\cdot 10^{-4}\mu \text{m}^{3} \text{s}^{-1}$, $\kappa_{d}=1 \text{s}^{-1}$ and $a_{\text{eff}}= 10^{-3}\mu\text{m}$.
The simulation was run for a time $t_{\text{sim}}= 10\text{s}$ and the copy numbers $N_{A}(t)$, $N_{B}(t)$, $N_{C}(t)$ over time $t$ were recorded. The used dimensionless time step is $\Delta \tau = \Delta t D_{\text{eff}} / a^{2}_{\text{eff}} = 10.1$.
Fig.~\ref{fig:rev_2d_3d} compares the simulation result with the numerical solution of the standard mass-action equations.

\section{Appendix}
The Green's function for free diffusion on a sphere with radius $R$ is given by \cite{brillinger1997a}
\begin{equation}\label{p_free_sphere}
p_{\text{free}}(\theta, t \vert \theta_{0}=0) = \frac{1}{4\pi}\sum_{l=0}^{\infty}(2l+1)\exp\bigg(-\frac{Dt l(l+1)}{R^{2}}\bigg)P_{l}(\cos \theta),
\end{equation}
where $P_{l}$ refers to the Legendre polynomials \cite{abramowitz1964handbook}.
 
Quite recently, an exact expression  has been obtained \cite{Grebenkov:2019aa} that gives the survival probability of an isolated pair undergoing a diffusion-influenced reversible reaction on the surface of a sphere.
Here, we use the GF rather than the survival probability; as pointed out in Ref.~\cite{Grebenkov:2019aa}, the corresponding expression can be readily inferred from the results presented in \cite{Grebenkov:2019aa}.
For the limiting case of a perfect sink one has \cite{chao1981localization, Grebenkov:2019aa}
\begin{equation}\label{p_abs_sphere}
p_{\text{abs}}(\theta, t \vert \theta_{0}) = \frac{1}{2\pi}\sum_{n=0}^{\infty}b^{2}_{n}\exp\bigg(-\frac{Dt \nu_{n}(\nu_{n}+1)}{R^{2}}\bigg)P_{\nu_{n}}(\cos \theta)P_{\nu_{n}}(\cos \theta_{0})
\end{equation} and $p_{\text{abs}}(\theta, t \vert \theta_{0})$ satisfies the completely absorbing BC
\begin{equation}
p_{\text{abs}}(\theta_{a}, t \vert \theta_{0})  = 0,
\end{equation}
where $\theta_{a}$ is given by 
\begin{equation}
\cos\theta_{a} = -\sqrt{1- (a/R)^{2}}.
\end{equation}
Here, $a, R$ denote the encounter radius of the isolated pair and the radius of the sphere, respectively.
Note that the Legendre functions of fractional order are given by Gauss' hypergeometric function $_{2}F(a,b;c;x)$, more precisely one has \cite{abramowitz1964handbook}
\begin{equation}
P_{\nu}(x) := \, _{2}F(-\nu, \nu+1; 1; (1-x)/2)).
\end{equation}
The $\nu_{n}$ are the solutions of the equation \cite{Grebenkov:2019aa}
\begin{equation}
P_{\nu}(\cos(\theta_{a})) = 0.
\end{equation}
Then, the $b_{n}$ terms can be calculated according to \cite{Grebenkov:2019aa}
\begin{equation}
b_{n} = \bigg(\int^{1}_{a} [P_{\nu_{n}}(x)]^{2}\,dx\bigg)^{-1/2}.
\end{equation}
To apply the expression given in Eq.~(\ref{p_abs_sphere}), we numerically calculated the first eight roots $\nu_{n}$ and  the corresponding coefficients $b_{n}$.

\section*{Acknowledgments}
This research was supported by the Intramural Research Program of the NIH, NIAID. 
\bibliographystyle{siam} 

\end{document}